\pdfoutput=1
\documentclass[11pt]{article}
\usepackage{amsfonts}
\usepackage{amsmath,amssymb,amsthm}
\usepackage{mathrsfs}
\usepackage{graphicx}

\usepackage[utf8]{inputenc}
\usepackage{caption}
\usepackage{subcaption}
\numberwithin{equation}{section}

\usepackage[sort&compress,numbers,colon,merge]{natbib}
\usepackage{multirow}
\usepackage{color}
\usepackage[colorlinks]{hyperref}
\hypersetup{
pdfauthor={Anibal D Medina},
pdftitle={Higgsino-like Dark Matter From Sneutrino Late Decays}
}

\usepackage{verbatim}
\usepackage{float}
\usepackage{tikz}
\usetikzlibrary{shapes,arrows}
\newcommand{\be}{\begin{equation}}
\newcommand{\ee}{\end{equation}}
\newcommand{\bea}{\begin{eqnarray}}
\newcommand{\eea}{\end{eqnarray}}

\def\4vol{{\int d^4x \sqrt{-g}}}

\def\beq{\begin{equation}}
\def\eeq{\end{equation}}
\def\bea{\begin{eqnarray}}
\def\eea{\end{eqnarray}}
\def\bitem{\begin{itemize}}
\def\eitem{\end{itemize}}

\newcommand{\nc}{\newcommand}

\nc{\nt}{\tilde{N}}
\nc{\ra}{\rightarrow}
\nc{\lsim}{\begin{array}{c}\,\sim\vspace{-21pt}\\< \end{array}}
\nc{\gsim}{\begin{array}{c}\sim\vspace{-21pt}\\> \end{array}}
\nc{\tnt}{\tilde{N}}
\nc{\tst}{\tilde{t}}

\nc{\LL}{L}
\nc{\vv}{\tilde{v}}

\title{}

\begin{document}
\allowdisplaybreaks[1]

\begin{titlepage}


\begin{center}
{\Large\textbf{Higgsino-like Dark Matter From Sneutrino Late Decays}}
\\[10mm]
{\large
Anibal D.~Medina$^{a,}$\footnote{\texttt{anibal.medina@cea.fr}} }
\\[5mm]
{\small\textit{$^a$ARC Centre of Excellence for Particle Physics at the Terascale, School of Physics, The University of Melbourne, Victoria 3010, Australia\\ and\\
Institut de Physique Th\'eorique, Universit\'e Paris Saclay, CNRS, CEA, F-91191 Gif-sur-Yvette, France}
}
\end{center}

\vspace*{1.0cm}
\date{\today}

\begin{abstract}
We consider Higgsino-like dark matter (DM) in the Minimal Supersymmetric Standard Model (MSSM) with additional right-handed
neutrino chiral superfields, and propose a new non-thermal way of generating the right amount of relic DM via sneutrino late decays.
Due to the large DM  annihilation cross-section, decays must occur at lower temperatures than the freeze-out temperature $T_d\ll T_{F,\tilde{\chi}^0_1}\sim \mu/25$,
implying a mostly right-handed lightest sneutrino with very small Yukawa interactions. In that context, the right amount of Higgsino-like DM relic density can be accounted for
if sneutrinos are produced via thermal freeze-in in the early Universe. 
\end{abstract}

\end{titlepage}

\setcounter{footnote}{0}


\section{Introduction}
\label{sec:Introduction}
Supersymmetry (SUSY) elegantly solves the quadratic ultra-violet (UV) sensitivity of the Higgs mass via the introduction of
particles (superpartners) with opposite statistics to each Standard Model (SM) particle. 
Stability of the proton naturally leads to the introduction of R-parity under which all superpartners are odd while the SM content is even.
Thus sparticles can only be created in pairs at colliders and the lightest supersymmetric particle (LSP) is stable, providing an interesting dark matter 
(DM) candidate. 

In the minimal supersymmetric extension of the SM a natural and well-studied DM candidate is the
lightest neutralino $\tilde{\chi}^0_1$, a linear combination of the wino $\tilde{W}$, bino $\tilde{B}$ and Higgsinos $\tilde{h}_u$ and $\tilde{h}_d$ superpartners. Given its weak couplings and for masses of order the EW scale, the lightest neutralino can provide the well-known "WIMP miracle" in
which the total amount of DM relic density is naturally obtained. Despite its appealing properties, neutralino DM in the MSSM is being pushed toward corners of parameter space, in particular due to the lack of positive signals in direct detection experiments that probe spin-independent (SI)~\cite{Akerib:2013tjd},~\cite{Aprile:2012nq} and spin-dependent (SD)~\cite{Aartsen:2012kia} scattering of DM particles off of nuclei target. Furthermore, the absence of discovery of superpartners at Large Hadron Collider (LHC)  and the discovery of a SM-like Higgs with a mass $m_h\sim 126$ GeV, seem to point towards a SUSY spectrum where at least part of the sparticle content have masses in the TeV range.  

Despite the increasing constraints on the sparticle masses and composition, a neutralino saturating the DM relic density and of almost pure Higgsino composition is able to evade current direct detection bounds due to its suppress coupling to the Higgs and Z-gauge boson~\cite{Cheung:2012qy}, see in particular Fig. 3 of Ref.~\cite{Cheung:2012qy}. Moreover, if the theory is to remain natural one expects the supersymmetric Higgs mass parameter $\mu\approx \mathcal{O}(100)$ GeV, making a neutralino LSP  with almost pure Higgsino composition $m_{\tilde{\chi}^0_1}\approx\mu$  a good candidate for DM.  As shown in Ref.~\cite{Huang:2017kdh}, for $\mu<0$ and in the case of a Higgsino-like neutralino with Bino admixture, there is a destructive interference in the diagrams that contribute to DM-nucleon scattering between the SM-like Higgs and  the non-standard Higgs H such that the latest LUX constraints can be avoided. Studies have shown that pure thermal Higgsino DM is under-abundant for masses below 1 TeV~\cite{Cirelli:2005uq}. This tension served as motivation for non-thermal ways of generating the right amount of Higgsino relic density~\cite{Acharya:2009zt},~\cite{Baer:2011hx}. 

In this work we propose an alternative non-thermal way of generating the right amount of Higgsino DM via late decays of sneutrinos. As has been well established by now, neutrinos are massive. A convenient manner of obtaining neutrino masses is through the addition of right-handed neutrinos to the SM content, which by means of  heavy Majorana masses leads to the type I see-saw mechanism of neutrino mass generation. When this extra content in the SM is supersymmetrized, we find that for the lightest sneutrino masses $m_{\tilde{\nu}}\gtrsim \mu$ and small Yukawa couplings $Y_N$ to the Higgsino-chargino sector, late decays of sneutrinos either directly to the lightest neutralino $\tilde{\chi}^0_1$ or cascading to it via decays to $\tilde{\chi}^0_2$ and $\tilde{\chi}^{\pm}_1$ can be efficient enough in generating the right amount of Higgsino DM relic density when sneutrinos are produced in the early Universe via decays of heavier SUSY particles (freeze-in scenario~\cite{Asaka:2005cn}, \cite{Hall:2009bx}).

The paper is organized as follows. In sec.~\ref{sec:HiggsinoDM} we briefly review the status of Higgsino DM in the MSSM and the current constraints on the parameter space. We move on in sec.~\ref{sec:relic} to describe the additon of the right-handed neutrino sector and how the decays of the lightest sneutrino can be efficient in generating the Higgsino DM relic density in a non-thermal way via late decays. Finally, our conclusions are given in sec.~\ref{sec:conclusion}.

\section{Higgsino dark matter}
\label{sec:HiggsinoDM}
The paradigm of Higgsino dark matter is well motivated both from arguments based on naturalness of the EW scale
as well as, on a more practical sense, from current collider constrains in SUSY searches at the LHC. In the large $\tan\beta\gg 1$ limit,
necessary to obtain the maximum value for the tree-level Higgs mass in the MSSM, $m_{h,tree}\approx m_Z$~\footnote{For this condition to hold a moderate value of $\tan\beta$ is necessary. One may be worried about possible flavour changing processes such as rare B-meson decays like $B^{0}_s \to \mu^{+}\mu^{-}$, whose SUSY contributions scale as $\tan^{6}\beta$. However, these contributions, which come from the exchange of the non-standard Higgses $H$ and $A$, are suppressed by $m^4_A$and thus become small in the decoupling limit, $m_A\gg m_h$~\cite{Buras:2002vd}.}, the usual measure of tuning~\cite{Barbieri:1987fn}, $\Delta =\max_i |d\log v^2/d\log \xi_i|\approx \mu^2+m^2_{H_u}$, where $\xi_i$ are the relevant parameters of the MSSM, implies that $\mu \lesssim $ few $\mathcal{O}(100)$ GeV  for a natural theory. Moreover, the recent discovery at the LHC  of a SM-like Higgs with mass $m_h\approx 126$ GeV implies in the MSSM that large radiative corrections are necessary to raise the tree-level Higgs mass. In these finite-loop corrections only third generation sparticles are relevant due to their coupling to the Higgs. Thus, an effective SUSY spectrum with only third generation sparticles, a Higgsino sector and all other sparticles decoupled becomes a natural option. On the other hand, the latest SUSY searches at the LHC~\cite{Aad:2014pda}~\cite{Chatrchyan:2013fea}, as well as flavour constraints from B-factories~\cite{Amhis:2012bh}, also highly constrain first and second generation sparticles as well as gluinos, pointing towards a natural spectrum.  We'd also like to point out that in the split versions of SUSY~\cite{ArkaniHamed:2004fb},~\cite{Arvanitaki:2012ps} where a natural EW theory is no longer a requirement,  a light Higgsino sector can arise due to chiral symmetry protection of the fermion masses, making Higgsino DM studies relevant for this case as well. \\
We concentrate  in subclasses of what are known as "Higgsino-world" scenarios~\cite{Kane:1998hz},~\cite{Baer:2011ec} in which squarks and sleptons of the MSSM have masses in the multi-TeV range, while $\mu$ is sub-TeV. In order to simplify our analysis, we take $ m_{\tilde{W}}\sim m_{\tilde{q}}\; , m_{\tilde{l}}$, decoupling the Wino from our effective theory.  Thus, we consider the range of masses: $|\mu|\ll m_{\tilde{B}} \ll  m_{\tilde{q}}\; , m_{\tilde{l}}\;, m_{\tilde{W}}$, with  light Higgsino-like charginos $\tilde{\chi}^{\pm}_1$ and two light Higgsino-like neutralinos $\tilde{\chi}^0_1$ and $\tilde{\chi}^0_2$. Though in order to get $m_h \approx 126$ GeV via a large trilinear $A_t$ the lightest stop could be sub-TeV, we assume for simplicity that the lightest stop has a mass above the TeV range. This kind of SUSY scenario has been thoroughly studied and it is well-known that  in the case of thermal production it leads to a very low relic density of neutralinos, in disagreement with the latest Planck results (at the 3$\sigma$ level)~\cite{Ade:2013zuv}: $0.1118 <\Omega_{DM} h^2<0.128$.  This is a consequence of the sizeable couplings involved and it implies that thermal Higgsino DM is under-abundant for $\mu \lesssim 1$ TeV~\cite{Cirelli:2005uq}.  Therefore, non-thermal ways of generating the correct amount of relic density have been proposed such as moduli field remnant from string theory decaying into a Higgsino-like neutralino LSP~\cite{Acharya:2009zt} or, in the midst of solving the strong-CP problem, a Peccei-Quinn axino annihilating to Higgsinos which provides a  Higgsino-dominated or axion dominated DM relic density (two species of DM) depending on the which type of annihilation dominates~\cite{Baer:2011hx}, among others~\cite{Fujii:2001xp, Seto:2005pj, Chun:2011zd}. In order to simplify our analysis, we take $ m_{\tilde{W}}\sim m_{\tilde{q}}\; , m_{\tilde{l}}$, decoupling the Wino from our effective theory.  It turns out that in this kind of SUSY spectrum, spin independent and spin dependent  direct detection constraints can be greatly ameliorated for an almost pure Higgsino DM due to its reduced coupling to the Higgs and the Z-gauge boson~\cite{Cheung:2012qy}. Indirect detection constraints from gamma ray observations at Fermi~\cite{Ackermann:2011wa} exclude non-thermal Higgsino-like DM for values of $|\mu|\lesssim 250$ GeV.  Therefore an acceptable region of MSSM parameter space which satisfies all relevant DM constraints is a non-thermal mostly Higgsino DM with $|\mu|\gtrsim 250$ GeV, $ M_1 \gg |\mu|$ and $\tan\beta\gg 1$, with the last constraint coming from the Higgs' mass requirements.


\section{Late decays of sneutrinos}
\label{sec:relic}

To generate the right amount of relic Higgsino-like DM with $m_{DM}\sim\mu\approx \mathcal{O}(100)$ GeV we need to resort to non-thermal ways. It has been well established by many experiments  that at least some of the neutrinos are massive and that the different flavours oscillate  in vacuum and matter. A simple way to generate neutrino masses in the SM is by adding at least 2 right-handed neutrino fields to the SM particle content. Given that this right-handed neutrinos are singlets under the SM gauge groups, a Majorana mass can be introduced for each of them which in conjunction with a Yukawa interaction involving the left-handed neutrino and the Higgs can be used in the well-known type I see-saw mechanism to generate small neutrino masses. Even in supersymmetric models we need to be able to account for massive neutrinos. In principle we could extend the MSSM to include 2 right-handed neutrino superfields and explain the solar and atmospheric neutrino mass differences. This however fixes the new Yukawa interactions between left-handed neutrinos and right-handed neutrinos and in practice does not allow late decays for the lightest snuetrino.  Thus we introduce 3 right-handed neutrino superfields $N_i$ with $i=a,b,c$ to the MSSM spectrum from which we can explain the solar and atmospheric neutrino mass differences by means of two of these.  The superpotential then takes the form $W=W_{MSSM}+M_{N_i} N_{i} N_{i}+y_{N_i} L.H_u N_i$, where $W_{MSSM}$ is the MSSM superpotential, $M_{N_i}$ the Majorana masses and $y_{N_i}$ the new Yukawa couplings.  Two of these Yukawa couplings are fixed by the atmospheric and solar mass differences and we use the third Yukawa interaction to generate the late out of equilibrium decays of the corresponding sneutrino.   We similarly introduce soft-breaking masses, bi-linear and tri-linear interactions for the scalar components, $\Delta\mathcal{L}_{soft}=-m^2_{\tilde{N}_i}|\tilde{N}_i|^2+((b_{N_i}/2) M_{N_i} \tilde{N}_i^2-A_{N_i} \tilde{L}.H_{u} \tilde{N}_i+h.c.)$, where we assume all couplings to be real.  Since we are interested in the DM picture, we decoupled from our low energy effective theory the sneutrinos corresponding to the solution of the solar and atmospheric mass differences (with indices $i=a,b$) by taking $m^2_{\tilde{L}_{a,b}}\sim m^2_{\tilde{N}_{a,b}} \gg m^2_{\tilde{N_c}}$.  Similarly, we assume for the Majorana masses $M_{N_a}\sim M_{N_b}\gg M_{N_c}$ decoupling the corresponding mostly right-handed neutrinos as well such that a see-saw with Yukawa couplings of $\mathcal{O}(1)$ is possible. Therefore, in effect we concentrate in a single superfield $N_c$ and in particular in its corresponding complex scalar component. Assuming CP-conservation in the sneutrino sector, we can decompose the chiral sneutrino fields as $\tilde{\nu}_L=(\tilde{\nu}_{L,1}+i\tilde{\nu}_{L,2})/\sqrt{2}$ and  $\tilde{N}=(\tilde{N}_{1}+i\tilde{N}_{2})/\sqrt{2}$, where from now on we understand $N=N_c$, and $L=L_c$ is the corresponding left-handed lepton flavour.  Then the sneutrino mass matrix reduces in the basis $(\tilde{\nu}_{L,1},\tilde{N}_{1} ,\tilde{\nu}_{L,2},\tilde{N}_{2} )$ to a block diagonal form,
\begin{eqnarray}
\label{CPevenMassMatrixTree}
{\small  \left(
\begin{array}{ccccc}
m^2_{LL} & m^2_{RL}+v \sin\beta Y_N M_N & 0 & 0\\
.  & m^2_{RR}-b_N M_N & 0   & 0\\
0 & 0 & m^2_{LL} & m^2_{RL}-v\sin\beta Y_N M_N \\
0 & 0 & . &  m^2_{RR}+b_N M_N
\end{array}
\right) }\nonumber
\end{eqnarray}
where $m^2_{LL}= m^2_{\tilde{L}}+Y^2_N v^2 \sin^2\beta+(m^2_Z/2)\cos^2 2\beta$, $m^2_{RR}=M_N^2+m^2_{\tilde{N}}+Y^2_N v^2\sin^2\beta$ and
$m^2_{RL}=-\mu Y_N v \cos\beta+v\sin\beta A_{N}$.  Denoting the mass eigenstates as $\tilde{\nu}'_i$ and $\tilde{N}'_i$ with $i=1,2$, we have that $\tilde{\nu}_i=\tilde{\nu}'_i\cos\theta_i - \tilde{N}'_i\sin\theta_i$ and $\tilde{N}_i=\tilde{\nu}'_i\sin\theta_i +\tilde{N}'_i\cos\theta_i $, where,
\begin{equation}
\tan 2\theta_i=\frac{2 (m^2_{RL}\pm v\sin\beta M_N Y_N)}{m^2_{LL}-(m^2_{RR}\mp b_N M_N)}\;.
\end{equation}

\begin{figure}[ht]
 \centering
\includegraphics[height=7cm]{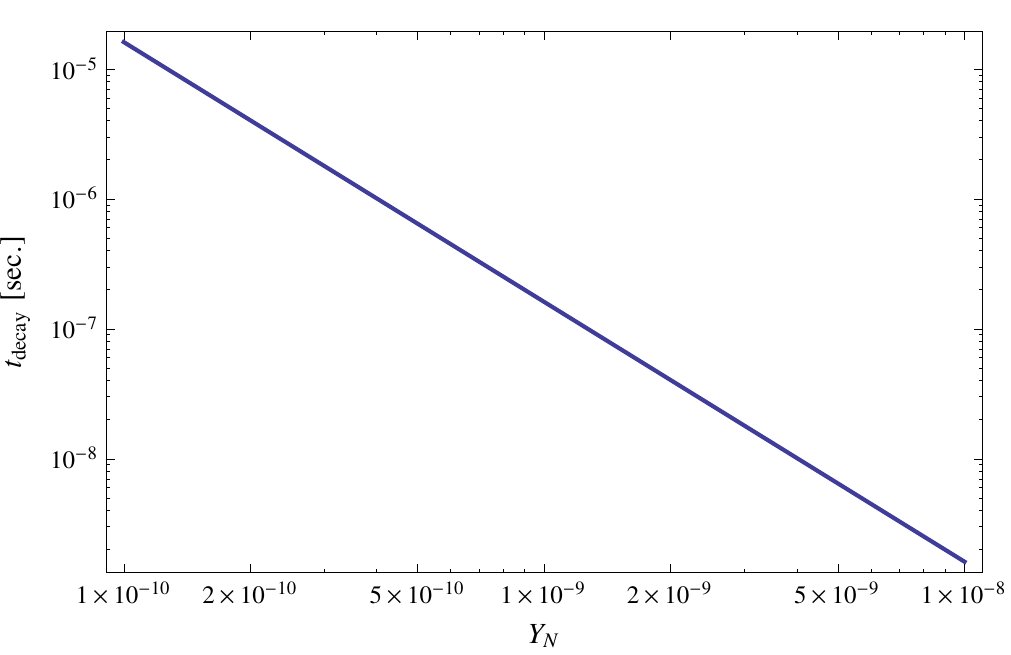}
\caption{ Decay time for $\tilde{\nu}_{0}\to \tilde{\chi}^{0}_1 \nu$ in sec. as a function of $Y_N$  for $m_{\tilde{\nu}_0}=500$ GeV and $m_{\tilde{\chi}^0_1}=300$ GeV. }
\label{fig:tdecay}
\end{figure}

At this stage we discuss the requirements for the non-thermal Higgsino-like DM generation from sneutrino late decays. We'll drop primes to make the notation less cumbersome. Calling the lightest sneutrino mass-eigenstate  $\tilde{\nu}_{0}$, we choose it  such that it corresponds to the CP-even sneutrino sector $i=1$~\footnote{We could have similarly chosen it to correspond to the CP-odd sector.}.  The decays that generate $\tilde{\chi}^0_1$ are: decays of $\tilde{\nu}_0$ into the lightest chargino, $\tilde{\nu}_{0}\to \tilde{\chi}^{\pm}_1 l^{\mp}\to f \bar{f}' l^{\mp}\tilde{\chi}^0_1$ where the chargino decay is via an off-shell charged $W^{\pm}$, and  decays  of $\tilde{\nu}_0$ into the second lightest neutralino $\tilde{\nu}_{0}\to \tilde{\chi}^{0}_2 \nu\to f \bar{f} \nu\tilde{\chi}^0_1$ where the $\tilde{\chi}^0_2$ decay is via an off-shell $Z$~\footnote{The decay $\tilde{\chi}^0_2\to W^{\pm *}\tilde{\chi}^{\pm}_1$ is kinematically suppressed}, and  finally direct decays  into the lightest neutralino  $\tilde{\nu}_{0}\to \tilde{\chi}^{0}_1 \nu$.  The neutralino-chargino (mostly Higgsino) sector states all have masses very close to $\mu$: $m_{\chi^{\pm}_{1}}\approx \mu$ and 
\begin{equation}
m_{\tilde{\chi}^{1,2}_{0}}\approx |\mu|+\frac{m^2_Z({\rm sign}(\mu) \pm\sin2\beta)\sin^2\theta_W}{2(\mu\mp M_1)}
\end{equation}
with ${\rm sign}(\mu)=\pm$ the sign of $\mu$, $M_1$ the Bino mass and $\theta_W$ the Weinberg angle.  Their composition is given by,
\begin{eqnarray}
\tilde{\chi}^{+}_{1}&=&\tilde{h}^{+}_u\;,\quad \tilde{\chi}^{-}_{1}=\tilde{h}^{-}_d\;, \nonumber\\
\tilde{\chi}^0_{1,2}&=&\frac{\tilde{h}^0_d\mp\tilde{h}^0_u}{\sqrt{2}}\pm\frac{|\sin\beta\pm\cos\beta|}{\sqrt{2}}\frac{m_Z}{M_1}\sin\theta_W \tilde{B} 
\end{eqnarray}
Even in the limit $M_{1}\to \infty$, radiative corrections to charginos and neutralinos can increase the mass splittings by hundreds of MeV~\cite{Bharucha:2012nx} and all DM considerations which we describe would still follow. However, in order to make the model more phenomenologically appealing in particular for collider searches we'll assume that $M_1\gtrsim |\mu|$. In the absence of CP-violation in the neutralino sector, due to the vector-like nature of the Bino coupling to sfermions, the partial decoupling of the Bino leads to the possible decay  $\tilde{\nu}_{0}\to \tilde{\chi}^0_1\; \nu$.  This decay goes through the left-handed sneutrino component of $\tilde{\nu}_0$, and thus via the coupling $g_1\sin\theta_1$. Similarly in the decay $\tilde{\nu}_{0}\to \tilde{\chi}^{-}_1 l^{+}$, the relevant coupling is proportional to $Y_L \sin\theta_1$, where $Y_L$ is the Yukawa coupling from the superpotential term $ Y_L H_d.L l_R$, with $l_R$ the right-handed charged lepton chiral superfield.  We then realize that the only possible way to suppress these latter type of decays via the left-handed sneutrino component of $\tilde{\nu}_0$ is demanding a very small mixing angle, $\sin\theta_1\ll 1$. 
Moreover, given the degeneracy of the neutralino-chargino sector, the final SM fermions should all be relatively light: at or below the di-tau threshold.  We calculated the 3-body decay $\tilde{\chi}^{\pm}_1\to f \bar{f}' \tilde{\chi}^0_1$ with $f$ and $f'$ light quarks (largest coupling kinematically available) and similarly $\tilde{\chi}^0_2\to \tilde{\chi}^0_1 \tau\bar{\tau}$, finding that these decays are instantaneous ($t_{3-body\;decay}\sim 10^{-14}\; s\ll t_{\tilde{\nu}_0\to\tilde{\chi}^0_1\nu}$) and implying that the late decays of sneutrinos $\tilde{\nu}_0$ in this scenario are governed effectively by either $Y_N$, $Y_L \sin\theta_1$ or $g_1 \sin\theta_1$.  We show in Fig.~\ref{fig:tdecay} the dependence of $t_{decay}$ as a function of $Y_N$ for fixed  $m_{\tilde{\nu}_0}=500$ GeV and $m_{\tilde{\chi}^0_1}=300$ GeV. 
Each $\tilde{\nu}_0$ decay generates one Higgsino-like DM particle $\tilde{\chi}^0_1$ which implies that the total relic density  $\Omega_{\tilde{\chi}^0_1}$ is simply  related to $\Omega_{\tilde{\nu}_0}$ by: $\Omega_{\tilde{\chi}^0_1}=(m_{\tilde{\chi}^0_1}/m_{\tilde{\nu}_0})\Omega_{\tilde{\nu}_0}+\Omega_{\tilde{\chi}^0_1,\;thermal}$, where $\Omega_{\tilde{\chi}^0_1,\;thermal}h^2\ll 0.1$.  The sneutrino late decays must happen after the LSP has frozen out, at temperatures of order $T_{F,\tilde{\chi}^0_1}\sim \mu/25$. However longer decay times are necessary since the decoupling of the Higgsino-like LSP is due to strong annihilation cross-section at freeze-out. If we were to instantly replenish the relic density from decays right after freeze-out it will annihilate once again. We demand that the LSP is effectively decoupled for number densities compatible with $\Omega_{\tilde{\chi}^0_1} h^2 \approx 0.1$. For that purpose we use the criterion that the annihilation rate of the LSP should be smaller than the Hubble expansion rate evaluated at the $\tilde{\nu}_0$-decay time. Assuming that no significant entropy is generated between the $\tilde{\nu}_0$ decays till nowadays and that the Universe is radiation dominated at the decay time epoch, the condition takes the form,
\begin{equation}
0.1 \frac{\rho_c}{m_{\tilde{\chi}^0_1}} \frac{s(x_d)}{s_0}\langle \sigma v (x_d)\rangle|_{\tilde{\chi}^0_1\tilde{\chi}^0_1\to SM} < 1.67 \sqrt{g_{*}}\frac{m^2_{\tilde{\chi}^0_1}}{M_{Pl}}\frac{1}{x^2_d}\label{outofeqdecay}
\end{equation}
where $x_d=m_{\tilde{\chi}^0_1}/T_d$ is related to the temperature at which the decays happen, $\langle \sigma v (x_d)\rangle|_{\tilde{\chi}^0_1\tilde{\chi}^0_1\to SM}$ is the annihilation rate of the LSP into SM particles, $\rho_c=8.06\times 10^{-47} h^2$ GeV$^4$ is the critical density of the Universe, $s_0 =2.22\times 10^{-38}$ GeV$^3$ is the entropy of the Universe today, $g_*$ are the active degrees of freedom and $s(x_d)=(2\pi^2/45) g_* (m^3_{\tilde{\chi}^0_1}/x_d^3)$  the entropy evaluated at the time of decay.  Taking the inequality  Eq.~(\ref{outofeqdecay}) and solving for $x_d$, we find that.
\begin{equation}
x_d \gtrsim \frac{2.6 \times 10^{-2} \sqrt{g_{*}} M_{Pl} \rho_c \langle \sigma v (x_d)\rangle|_{\tilde{\chi}^0_1\tilde{\chi}^0_1\to SM}}{s_0}\;.\label{xdEq}
\end{equation}
Typical thermal relic densities for a Higgsino-like LSP are of the order of $\Omega_{\tilde{\chi}^0_1,\;thermal}h^2\sim 10^{-2}$, see for example~\cite{Cirelli:2005uq}, \cite{Cirelli:2007xd}. Since the Higgsino annihilations are s-wave dominated, we can get an estimate of the corresponding thermally averaged annihilation cross-sections  $\langle \sigma v (x_d)\rangle|_{\tilde{\chi}^0_1\tilde{\chi}^0_1\to SM}\sim 2\times 10^{-8}$ GeV$^{-2}$ and therefore on $x_d$ via  Eq.~(\ref{xdEq}), $x_d \gtrsim 230$.  We see that for values of $m_{\tilde{\chi}^0_1}\sim \mu \gtrsim 300$ GeV, this corresponds to a decay temperature $T_d \lesssim 1$ GeV or similarly to a decay time $t_d\gtrsim 10^{-6}$~s, safely below Big Bang Nucleosynthesis (BBN) times. \\
	The demand of a sufficiently late decay of sneutrinos $\tilde{\nu}_0$ implies in particular that for $m_{\tilde{\nu}_0}\gtrsim \mu$, $ Y_N \lesssim 10^{-10}$  as can be seen from Fig.~\ref{fig:tdecay}, and that $\sin\theta_1 \lesssim 5\times 10^{-8}$, the latter constraint coming from $\tilde{\nu}_0$ decays via the LSP Bino component~\footnote{Even in the limit $M_1\to \infty$, the mixing angle $\sin\theta_1\lesssim 10^{-5}$ from decays mediated via $Y_L$.}.  Thus, the associated light neutrino $\nu$ is basically massless, while the heavy neutrino $\nu_H $ is mostly right-handed with a mass $m_{\nu_{H}}=M_N$.
The sneutrino $\tilde{\nu}_0$ is also highly right-handed and interacts minimally with early Universe plasma. Self-annihilations and possible co-annihilations cross-sections are all too small to reproduce the correct $\tilde{\chi}^0_1$-relic density, suppressed by either $\sin^2\theta_1$, $\sin^4\theta_1$, $Y_N^2$ or combinations of these, see Ref.\cite{Gopalakrishna:2006kr}. Entropy generation in the decays is minimal since the Universe is radiation dominated and thus decays are not relevant in reducing the DM relic density. We conclude that the demand of sufficient late decays for the sneutrinos and the appropriate value of the $\tilde{\chi}^0_1$ relic density moves us to consider a model where the sneutrinos $\tilde{\nu}_0$  density is generated via decays of heavier SUSY particles in what is known as an example of the freeze-in mechanism~\cite{Hall:2009bx},~\cite{Petraki:2007gq},~\cite{Medina:2011qc}. 

These decays of heavier SUSY particles can also produce the associated right-handed neutrino $\nu_H$, which depending on the value of its mass has the potential of changing the LSP relic density. Given the interaction terms on the superpotential, the decays controlled via the Yukawa coupling $Y_N$, $\tilde{l}^{+}\to \tilde{\chi}^+_1 \nu_H$ and $\tilde{\nu}_1\to \tilde{\chi}^0_1 \nu_H$, are in principle the main $\nu_H$-production channels if kinematically available. Due to the gauge interactions of the mostly left-handed sparticles $\tilde{l}$ and $\tilde{\nu}_1$, they are in kinetic and thermal equilibrium with the early Universe plasma. Notice that both $\tilde{\chi}^0_1 $ coming from  $\tilde{\nu}_1\to \tilde{\chi}^0_1 \nu_H$ and the one coming from the chargino decay thermalize with the plasma and become part of the thermal relic density of neutralinos, which we know to be suppressed due to the Higgsino interactions. If $M_N < m_{\tilde{\nu}_0}+m_{\tilde{\chi}^0_1}$, then $\nu_H$ cannot possibly decay into a pair of SUSY particles and thus its decay is always to SM particles as in the case $\nu_H\to \nu h$, if kinematically available.  If $m_{\tilde{\nu}_0}+m_{\tilde{\chi}^0_1}<M_N<m_{\tilde{\nu}_1}+m_{\tilde{\chi}^0_1}$, then $\nu_H$ could decay into the NLSP and the LSP. However, the ratio of branching fraction for such decay with respect to the decay into a Higgs and a neutrino goes as $BR(\nu_H\to \tilde{\nu}_0 \tilde{\chi}^0_1)/BR(\nu_H\to \nu h)\approx \sin^2\theta_1\times (1-m^2_{\tilde{\nu}_0}/M^2_N)^2$, highly suppressed by the small mixing angle. Thus, in both of these kinematical regimes, we expect to have late decays of the heavy mostly right handed neutrino $\nu_H$ into SM particles which are governed by the coupling $Y_N$ and that for the values of $Y_N$ considered will not disrupt successful BBN.

 We take $m^2_{LL}\gg m^2_{RR}$ and $m^2_{RL}\ll m^2_{LL}$ which implies that $\sin\theta_1\ll 1$, making the mostly left-handed sneutrinos heavy $m_{\tilde{\nu}_{1,2}}\gg m_{\tilde{\nu}_{0}}$. Furthermore we appropriately choose $b_N$ such that $m_{\tilde{N}_2}>m_{\tilde{N}_1}\approx m_{\tilde{\nu}_0}$. The relevant decays to produce $\tilde{\nu}_0$-sneutrinos are $\tilde{\nu}_1\to \tilde{\nu}_0 h$, $\tilde{l}^{\pm}\to\tilde{\nu}_0 W^{\pm}$  and $\tilde{\nu}_1\to \tilde{\nu}_0 Z$, since these are the only sparticles with which $\tilde{\nu}_0$ couples to.  We can solve for the "would-be" relic density of $\tilde{\nu}_0$ integrating the Boltzmann equation from the re-heating epoch ($T_R$) till the time when the reaction rate decouples ($T_F$) to find,
\begin{equation}
\Omega_{\tilde{\nu}_0}=\sum_{i=\tilde{\nu}_1,\tilde{l}^{\pm}}\frac{m_{\tilde{\nu}_0}}{\rho_c} \frac{s_0}{s(x_{F,i})}\frac{1}{x_{F,i}^3}\frac{1}{H(m_i)}\int^{x_{F,i}}_{x_{i,R}} dx_i x_i^4 C[x_i]\label{decayrelic}
\end{equation}
where $H(m)=1.67\sqrt{g_*}m^2/M_{Pl}$ and $C[x]$ is the collision term, which for decays takes the form,
\begin{equation}
C[x]=\int e^{-\frac{E_i}{T}}\Gamma_i \frac{d^3p_{i}}{(2\pi)^3}\label{collisiontermdec}
\end{equation}
with $\Gamma_i=\Gamma^{CM}_i/\gamma_i$, the decay rate related to the center of mass (CM) decay rate via the time dilation factor $\gamma_i=E_i/m_i$.
In the case at hand, where $\Gamma^{CM}_i$ is a constant, we can readily do the integral  Eq.~(\ref{collisiontermdec}) and obtain $C[x_i]=m_i^3 \Gamma^{CM}_i K_1(x_i)/(2\pi^2 x_i)$, where $K_1(x_i)$ is the modified Bessel function of the second kind. For $T_R\gg m_{\tilde{\nu}_1}, m_{\tilde{l}^{\pm}}$, we get an approximate solution for the integral,
\begin{equation}
\int^{x_{F,i}}_{x_{R,i}} dx_i x_i^4 \frac{K_1(x_i)}{x_i}\approx \frac{3}{2}\pi-\frac{m_i^3}{3T^3_R}
\end{equation}
which shows that the contributions from decays is insensitive to the re-heating temperature as long as it is large enough.
The expressions for the decays $\tilde{\nu}_1\to\tilde{\nu}_0 h$  and $\tilde{l}^{\pm}\to\tilde{\nu}_0 W^{\pm}$  in the CM frame are,
\begin{eqnarray}
\Gamma_{\tilde{\nu}_1\to\tilde{\nu}_0 h}&=&\frac{|C_{\tilde{\nu}_1\tilde{\nu}_0 h}|^2}{16 \pi m_{\tilde{\nu}_1}}\sqrt{1-2\frac{(m^2_h+m^2_{\tilde{\nu}_0})}{m^2_{\tilde{\nu}_1}}+\frac{(m^2_{\tilde{\nu}_0}-m_h^2)^2}{m^4_{\tilde{\nu}_1}}}\nonumber\\
\Gamma_{\tilde{e}^{\pm}\to\tilde{\nu}_0 W^{\pm}}&=&\frac{g^2\sin^2\theta_1 }{32\pi m_{\tilde{e}^{\pm}}}\left(-(2m^2_{\tilde{e}^{\pm}}+2m^2_{\tilde{\nu}_0}-m^2_{W})\right.\nonumber\\
&+&\left.\frac{1}{4m^2_{W}}(2m^2_{\tilde{e}^{\pm}}-2m^2_{\tilde{\nu}_0})^2\right)\sqrt{1-\frac{m^2_{\tilde{\nu}_0}+m^2_W}{ m^2_{\tilde{e}^{\pm}}}}\nonumber\\
\end{eqnarray}
 where $C_{\tilde{\nu}_1\tilde{\nu}_0 h}\approx A_N (\cos^2\theta_1-\sin^2\theta_1)$ is the $\tilde{\nu}_1\tilde{\nu}_0 h$-coupling.  Similarly $\Gamma_{\tilde{\nu}_1\to\tilde{\nu}_0 Z}=\Gamma_{\tilde{e}^{\pm}\to\tilde{\nu}_0 W^{\pm}}/2$ with the replacements $m_{\tilde{e}^{\pm}} \to m_{\tilde{\nu}_1}$, $m_W \to m_Z$. Plugging these expressions into Eqs.~(\ref{decayrelic}, \ref{collisiontermdec}) and calculating the relic density of $\tilde{\chi}^0_1$  assuming $m_{\tilde{l}^{\pm}}=m_{\tilde{\nu}_1}$, we find that in order to get a DM relic density $\Omega_{\tilde{\chi}^0_1} h^2 \sim 0.1$,  we need $A_N \sim  10^{-9}$ GeV, independently of $\tan\beta$ and only mildly dependent on $m_{\tilde{\nu}_1}$, $m_{\tilde{\nu}_0}$  and $m_{\tilde{\chi}^0_1}$. This at the same time implies that $\sin\theta_1 \lesssim 10^{-12}$. We work  in a bottom-up approach where we take the parameters at low-energies, in particular $A_N$,  to include already any possible running effects~\footnote{Nonetheless, it is possible to envision cancellations taking place in the running of the trilinear coupling $A_N$ between the different contributions contributions as to keep its value small at low energies, see Ref.~\cite{Hisano:1995cp}}. We do not worry about possible UV-completions, hidden sector dynamics or transmission of SUSY breaking to the visible sector,  given that we are working from a low-energy point of view and we are interested in the low-energy phenomenology. 
 
\begin{figure}[ht]
 \centering
\includegraphics[height=7cm]{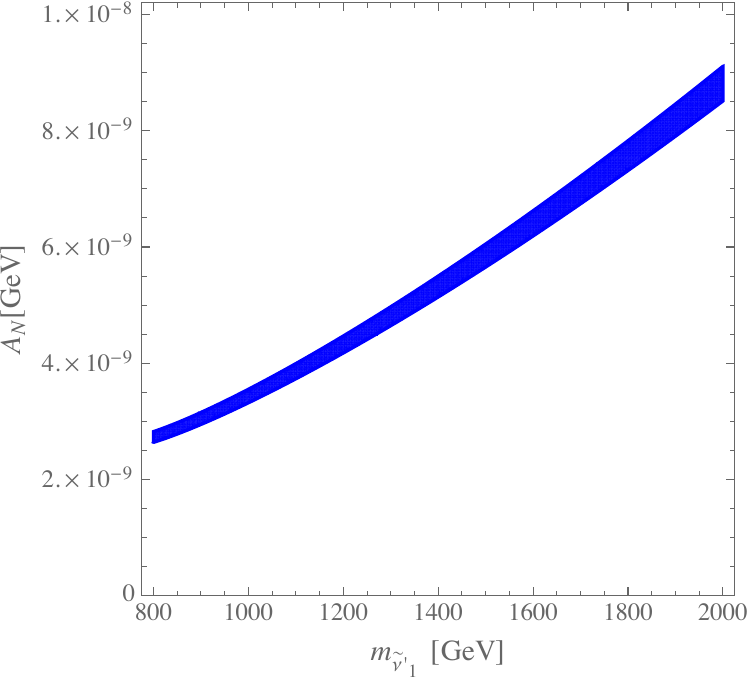}
\caption{ Region of $m_{\tilde{\nu}_1}$ [GeV] vs $A_N$ [GeV] where $0.1118 <\Omega_{\tilde{\chi}^0_1} h^2<0.128$ in agreement with Planck,
with $m_{\tilde{\chi}^0_1}=300$ GeV, $m_{\tilde{\nu}_0}=500$ GeV and $m_{\tilde{l}^{\pm}}=m_{\tilde{\nu}_1}$. }
\label{fig:relic}
\end{figure}

  
  We show in Fig.~\ref{fig:relic} a typical region of parameter space where the correct DM relic density is obtained.

Given the smallness of the interactions for the light sneutrino, we expect that the collider phenomenology of this scenario to be very similar to the "Higgsino-world" scenarios. 
In regards to Leptogenesis, though it goes beyond the scope of this work, it may still be possible to accommodate with the Majorana states involved in the generation of the heavier active neutrino masses $m_{\nu_2}$ , $m_{\nu_3}$. 


%

\section{Conclusion}
\label{sec:conclusion}

Motivated by the type I see-saw mechanism for the generation of neutrino masses,  we have shown an alternative non-thermal way of generating Higgsino-like DM in a trivial extension of MSSM via late decays of highly sterile, mostly right-handed sneutrino $\tilde{\nu}_0$. Due to the smallness of the $\tilde{\nu}_0$-interactions, in order not to overclose the energy density of our Universe, we are force to consider that the sneutrino $\tilde{\nu}_0$ is never in thermal equilibrium with the early Universe plasma, and thus its number density is produced via the freeze-in process in the form of decays of heavier SUSY particles, allowing a correct Higgsino-like DM relic density to be obtained in accordance with the latest Planck measurements.


\section*{Acknowledgements}
We thank Michael A.~Schmidt, Timothy Trott and Carlos E. M. Wagner for helpful discussions
and comments. During part of this work ADM was supported by the Australian Research Council. This work is supported by the European Research Council (ERC) Advanced Grant Higgs@LHC.

\bibliography{higgsinoLSPsneuNLSP}


\end{document}